\documentclass[12pt]{iopart}
\input epsf
\include{psfig}

\begin{document}

\title{Mode-Coupling Theory of Colloids with Short-range Attractions}

\author{K\ A\ Dawson$^\dag$, G\ Foffi$^\dag$,
F\ Sciortino$^\ddag$, P\ Tartaglia$^\ddag$ \\ and E\ Zaccarelli$^\dag$}
\address{$^\dag$Irish Centre for Colloid Science and Biomaterials,
Department of Chemistry, University College Dublin, Belfield, Dublin 4,
Ireland}
\address{$^\ddag$Dipartimento di Fisica and Istituto Nazionale per la Fisica
della Materia, Universit\`{a} di Roma {\it La Sapienza}, P.le A.
Moro 2, I-00185 Roma, Italy}


\begin{abstract}
Within the framework of the mode-coupling theory of super-cooled
liquids, we investigate new phenomena in colloidal systems on approach
to their glass transitions.  When the inter-particle potential
contains an attractive part, besides the usual repulsive hard core,
two intersecting liquid-glass transition lines appear, one of which
extends to low densities, while the other one, at high densities,
shows a re-entrant behaviour. In the glassy region a new
type of transition appears between two different types of
glasses. The complex phenomenology can be described in terms of higher
order glass transition singularities. The various glass phases are
characterised by means of their viscoelastic properties.
The glass driven by attractions  has been associated
to particle gels, and
the other glass is the well known repulsive colloidal glass. These
correspondences, in associations with the new predictions of glassy
behaviour mean that such phenomena may be expected in colloidal systems
with, for example, strong depletion or other short-ranged attractive
potentials.
\end{abstract}

\submitto{JPCM}
 
\maketitle
                
\section{Introduction}
\label{sec:intro}
The experimental study of the glass transition in colloidal systems
\cite{Pusey91} has been a very important test case to assess the
validity of theories concerning the formation of an amorphous solid
from super-cooled liquids. In particular the mode-coupling theory
(MCT) has been useful when applied to colloids, modeled as spherical
particles interacting through a hard sphere purely repulsive
potential~\cite{Goetze91}.  In this case the MCT predicts the existence
of a critical volume fraction $\phi$ where the system undergoes an
ergodic-nonergodic transition, which was observed experimentally using
quasi-elastic dynamic light scattering~\cite{Megen91}.  From the
physical point of view the MCT describes in a fairly accurate way the
so-called cage effect, i.e. the fact that at high densities molecular
motions of particles are constrained by the presence of the
surrounding ones which form a cage around each particle.  Prior to
reaching the MCT transition, we may think of the particle
vibrating within their cages at short time-scales, and escaping from
their cages at somewhat longer time-scales.  The two time scales show
up in a well-defined way, when the liquid system gets closer to the
critical threshold, as distinct relaxation regimes
separated by a plateau region which
becomes more and more extended as criticality is approached.
Close and above the plateau, there is a
power law and the plateau ends with another power law prior to enter
into the so-called full $\alpha$ regime.
The power-law region is called $\beta$ correlator in the framework
of $MCT$.
The $\alpha$ decay is quite well
phenomenologically described by a stretched exponential.

We have briefly sketched the main features of the MCT for colloids
considered as hard spheres and proceed to consider the novel
consequences of adding an attractive contribution to the
inter-particle potential.  In this paper we mainly consider the
effects of using a hard core followed by a square well potential in
order to mimic the interactions between colloidal particles.  In real
systems this can be obtained, for example, by covering the surface of
the particles with a polymer coating or with depletion
interaction~\cite{Poon93}.  A description of such systems using the MCT
approach has revealed the existence of a set of new and interesting
phenomena, that we briefly summarise in what
follows~\cite{Fabbian99,Dawson00}.
In the temperature $T$ and volume fraction
$\phi$ plane MCT predicts two lines of transition from liquid to
glass.One of the lines extends to high temperatures, it can be traced
essentially to the repulsive part of the potential and tends
asymptotically for high temperatures to the value $\phi$ corresponding
to the critical value for hard spheres. We will call it the repulsive
glass transition line (RGL).  On lowering the temperature the
transition line moves toward higher values of $\phi$ and gives rise
to a re-entrant behaviour, i.e. one can pass from a liquid to a glassy
state either by lowering or by raising $T$.  In other words the liquid
phase tends to exist in a region which extends more deeply into the
glassy phase compared to the hard spheres case.  The origin of this
unusual effect is that if the interaction is short ranged enough,
there can, in this temperature and density regime, be partial
cancellation of the repulsive and attractive interactions leading to a
phenomenon for glasses, not unlike that of a theta point for polymers.

The other glass transition line, originating in a relatively
well-defined energy scale of the well-depth, is almost parallel to the
$\phi$ axis and extends on one side of the binodal to the other, until
it eventually crosses the repulsive transition line. We call it the
attractive glass transition line (AGL). For sufficiently narrow
well-widths and on the high $\phi$ end the AGL enters the glass region
and terminates in an endpoint. Thus the two sides of the line separate
two different types of glassy system that we will characterise through
their mechanical properties.  The endpoint of the AGL is a higher
order glass transition point (called $A_3$), which will be shown to
have special dynamical properties, namely the relaxation slows down
dramatically on approaching it.  Upon increasing the width of the
potential well, the length of the AGL shrinks and its extension into
the glass region reduces to zero, and as a consequence the endpoint
turns into a higher transition point (called $A_4$).  We will
characterise this rather complex behaviour of the system, where
repulsion and attraction compete, using two typical dynamical
quantities, the shear viscosity, which characterises the liquid phase,
and the elastic shear modulus for the amorphous glass.

Aspects of the MCT have been tested in many different cases, both
experimentally and with computer molecular dynamics \cite{Goetze99}.
The most extensive and accurate experimental check has been performed
in the liquid-glass transition of colloidal systems, treated as hard
sphere systems, and studied with dynamic light scattering
\cite{Megen91}. The agreement with MCT is quantitatively satisfactory
\cite{Megen95}.  In a similar fashion computer simulations have been
used to study the glass transition in simple model systems, as diverse
as Lennard-Jones binary systems \cite{Kob95}, the SPC/E model for
water \cite{Gallo96}, orthoterphenil \cite{rinaldi} and silica \cite{kob}.
In all these cases evidence shows that these systems
undergo a kinetic glass transition, and the molecular dynamics
is well accounted for by the idealised MCT of super-cooled liquids.
On the other hand there is extensive evidence, mainly due to
experimental results in colloidal systems, that cannot be interpreted
in terms only of hard-core potentials.

Dense systems of colloidal particles characterised by a hard core and
strong short-ranged attractions have been realized experimentally by
adding polymers to either a suspension of colloidal hard spheres
\cite{Poon93}, in solutions of sterically stabilised particles when
decreasing the solvent quality \cite{Verduin95}, and in copolymer
micellar systems when changing the temperature \cite{Lobry99}.  Such
systems were also studied by Monte Carlo simulations
\cite{Kranendonk88}.  The facts that cannot be simply explained using
hard-core potentials are the following.

(i) An amorphous material can be formed by increasing the attraction
strength
even though the volume fraction is kept well below the value of the
hard sphere glass transition
\cite{Verduin95,Meller99,Grant93,Rueb97,Rueb98}.

(ii) In mixtures of colloids and polymers, melting of the glass states
is obtained by increasing the strength of a short range attraction
by the addition of small polymers \cite{Poon93}.

(iii) Using solvents of decreasing quality \cite{Verduin95} in
polymer coated colloidal particles the long time limit of the density
time correlation function at small
wave vectors are much larger than in hard-sphere systems.

(iv) Viscoelastic measurements for intermediate frequencies found
strongly concentration dependent elastic moduli
\cite{Meller99,Grant93,Rueb97,Rueb98}.

(v) A system in which an anomalous dynamical behaviour has been reported is
a polymer solution (called L64) where structural arrest accompanied by a
logarithmic-like decay of the density correlator is observed
\cite{Mallamace00}.

The paper is organised as follows.  In Section~\ref{sec:Sofq} we
report the details of the calculation of the structure factor for a
square-well system (SWS), the only input needed for the mode-coupling
dynamics, which is described in Section~\ref{sec:MC}.  The resulting
phase diagram is described in Section~\ref{sec:phase}, the time
dependent density correlation functions in Section~\ref{sec:dyn} and
finally the various phases are characterised by their mechanical
properties in Section~\ref{sec:mech}. In Section~\ref{sec:conc} we
report our conclusions.

%
\section{Calculation of the structure factor}   %
\label{sec:Sofq}                                %
%
In the framework of the MCT the only quantity that is needed in order
to calculate the dynamical properties of a super-cooled liquid
approaching the glass transition is the structure factor.  It is well
known that, given an inter-particle potential satisfying the
Ornstein-Zernike equation for the space-dependent pair correlation
function $g(r)$, one needs a closure approximation in order to solve
the equation. Many approximations of this type have been proposed and
solved in the case of simple model potential \cite{Hansen86}.  In the
case of the square-well potential the
Wiener-Hopf method as formulated by Baxter \cite{Baxter68} can be used.
The
interaction potential $V(r)$ for particles a distance $r$ apart, is
obtained with a hard-core repulsion for $r<d$, and the negative
attractive value $-u_0$ within the range $d<r<sD$.

The structure factor is specified by three control parameters, i.e.
the packing fraction $\phi$ related to the hard cores, the temperature
$T$, and the relative width $\epsilon = 1 -d/D$ of the attraction
shell.  In the Baxter approach the structure factor $S_q$ is expressed
in terms of the Fourier transform $\widetilde{Q}(q)$ of the factor
function $Q(r)$, a continuous real function defined for $r\ge 0$,

\begin{eqnarray}\label{eq:Sq-a}
S_q^{-1}   &=& \widetilde{Q}(q)\widetilde{Q}^*(q)\,,\\\label{eq:Sq-b}
\widetilde{Q}(q) &=& 1 - 2 \pi \rho \int_0^\infty {d}r \exp(i q r) Q(r)\,.
\end{eqnarray}
%

The funcion $Q(r)$ is related to the direct correlation function
$c(r)$ \cite{Hansen86} and both functionss vainsh beyond the distance
$D$. For $0 \le r \le D$, the following equation for $Q(r)$ holds

\begin{equation}\label{eq:WH1-b}
r c(r) = - Q'(r) + 2 \pi \rho \int_r^R {d}s\, Q'(s)Q(s-r)\,.
\end{equation}
In addition for $r>0$

\begin{equation}\label{eq:WH2-a}
r h(r) = - Q'(r) + 2 \pi \rho \int_0^R {d}s (r-s) h(|r-s|) Q(s)\,.
\end{equation}
For the SWS, $g(r)=0$ is fulfilled for $0<r<d$, and therefore, using
$h(r)=g(r)-1$, equation~(\ref{eq:WH2-a}) splits into three
sub-equations.  The result for the middle part, $\Delta \le r \le d$,
where $\Delta=D-d$, is very simple since the formula known from the
theory for the hard-sphere system is reproduced

\begin{equation}\label{eq:WHQ-a}
Q'(r) = a r + b\,,
\end{equation}
where the coefficients $a$ and $b$ are introduced by
\begin{equation}\label{eq:WH:ab}
a = 1 - 2 \pi \rho \int_0^{d+\Delta} {d}s \,Q(s)\,,
\quad
b=2 \pi \rho \int_0^{d+\Delta} {d}s \,s\, Q(s)\,.
\end{equation}
Defining $G(r)=rg(r)$ one finds, for small distances,
$0\le r \le \Delta$, where $\Delta=D-d$,

\begin{equation}\label{eq:WHQ-b}
Q'(r) = a r + b - 2\pi\rho \int_{r+d}^{d+\Delta}{d}s\,G(s-r)Q(s)\,,
\end{equation}
and for the attraction shell, $d\le r\le D$, one obtains

\begin{equation}
\label{eq:WHQ-c}
Q'(r) = a r + b -G(r) + 2\pi\rho \int_{0}^{r-d}{d}s\,G(r-s)Q(s)\,.
\end{equation}

\begin{figure}[htbp]
\begin{center}
\mbox{\psfig{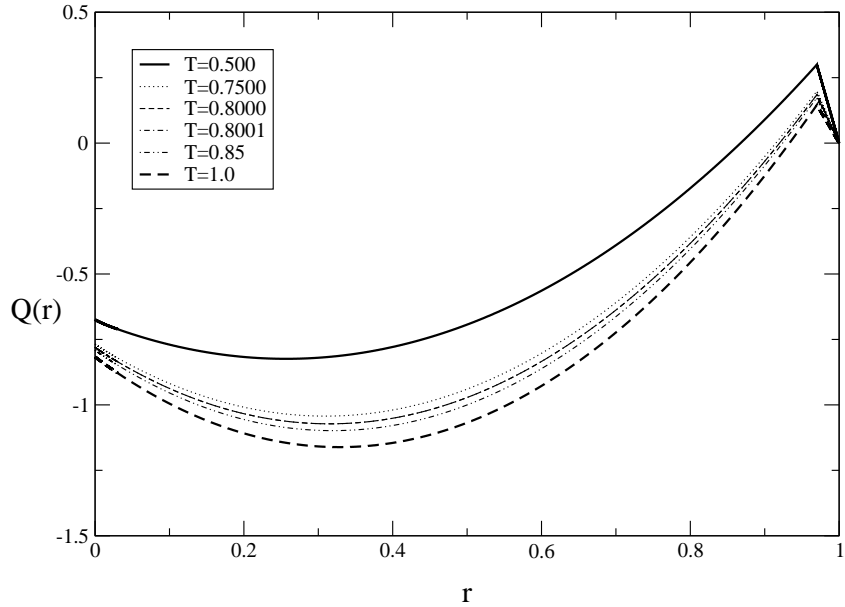}}
\end{center}
\caption{The Baxter factor function $Q(r)$ at $\epsilon=0.03$,
$\phi=0.504449$ and different
temperatures. The hard-core diameter is chosen as the unit of length.
}
\label{fig:fig1}
\end{figure}
\begin{figure}
\begin{center}
\mbox{\psfig{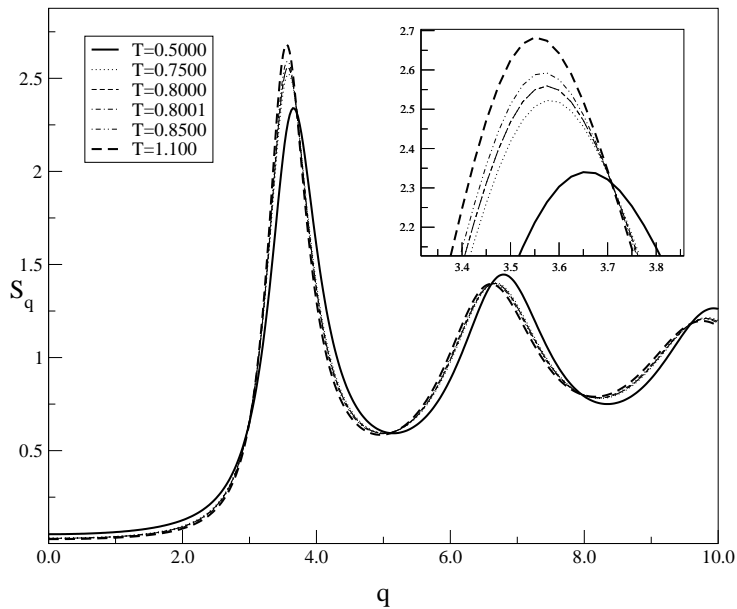}  }
\end{center}
\caption{The structure factors function in the PY approximation at various
temperatures, $\epsilon=0.03$ and $\phi=0.504449$. The inset shows the
maxima of the
structure factors.}
\label{fig:fig2}
\end{figure}

A closure approximation for $c(r)$ must be introduced into
equation~(\ref{eq:WH1-b})
in order to complete the system of equations~(\ref{eq:WH1-b}) and
(\ref{eq:WHQ-c}).
Using the Percus-Yevick (PY) approximation~\cite{Hansen86},
according to which

\begin{equation}
c(r) = g(r) \left[1-exp\left({{V(r)}\over{k_B T}}\right)\right]
\end{equation}
inside the hard core in equation~(\ref{eq:WH1-b}) and
equation~(\ref{eq:WHQ-c}) leads to the approximation valid for
$d \le r \le d+\Delta$

\begin{eqnarray}
\label{eq:PYA}
e^{-u_0/k_B T} G(r) =&& a r + b - 2\pi\rho\int_r^{d+\Delta}{d}s\,
                          Q'(s) Q(s-r)\nonumber\\
                &&+ 2\pi\rho\int_0^{r-d}{d}s\,G(r-s) Q(s)\,.
\end{eqnarray}
Equations~(\ref{eq:WHQ-c}) and (\ref{eq:PYA})
for $Q(r)$ and $G(r)$ are solved numerically choosing
on each of the three intervals of the variable
a grid of equally spaced points $r_n$, where $n=1,2,\dots,1000$.
The integral in equation~(\ref{eq:Sq-b}) is then performed
to obtain $\widetilde{Q}(q)$ and hence $S_q$.
The typical values for the factor function $Q(r)$ are shown in
figure~\ref{fig:fig1}, while figure~\ref{fig:fig2} shows the
corresponding structure factors $S_q$ for various temperatures,
given in term of the well depth $u_0$ as $k_b T / u_0$.

In a first attempt to solve the problem associated with the attractive part
of the inter-particle potential the Baxter potential has been
used \cite{Baxter68}, i.e.
the limit of a square-well in which the range of the potential vanishes
while its depth increases in such a way that their product remains
constant \cite{Fabbian99}. The mean spherical approximation has also
been used in conjunction with an attractive Yukawa potential in an attempt
to
explain colloidal gelation \cite{fuchs}.
We may note that more recently other methods of generating
the structure factors have been explored \cite{Dawson00,unpublished}.
The main results reported here are found not to depend on these details,
and we therefore here only comment on the square well potential.

%
\section{The mode-coupling equations}  %
\label{sec:MC}                         %
%
The MCT of the ideal liquid-glass transition is capable of describing
the ergodic to non-ergodic transition in the normalised density
correlation function

\begin{equation}
\phi_q(t)={{\langle \rho_{{\vec{q}}}^{*}(t)\rho_{\vec{q}}\rangle}
\over
{\langle|\rho_{\vec{q}}|^2\rangle}}.
\end{equation}
In the long time limit $\phi_q(t)$ shows a discontinuous variation
from a vanishing value in a super-cooled liquid to a finite value
different from zero $f_q$ in a glassy state, on varying the external
control parameters. In the latter state the system experiences a
structural arrest.  In real systems the transition point is never
reached since new dynamical mechanisms set in which bypass the
ergodicity breakdown.
The MCT equations are

\begin{equation}
\label{eq:MCT}
\ddot{\phi_{q}}(t)+\Omega_{q}^{2}\phi_{q}(t)+\nu_{q}\dot{\phi_{q}}(t)+
\Omega_{q}^{2}\int_{0}^{t}m_{q}(t-t')\dot{\phi_{q}}(t')dt'=0
\end{equation}
where the mode-coupling memory functional $m_q$ is given by

\begin{equation}
\label{eq:functional-a}
m_q(f)  = \frac{1}{2} \int{ {{{d}^3k} \over
(2 \pi )^3} V_{\vec{q},\vec{k}} f_k f_{|\vec{q}-\vec{k}|}}\,.
\end{equation}
The mode-coupling vertices are determined by the density $\rho$, the
structure factor $S_q$ and
the direct correlation function $c_q=(1-1/S_q)/\rho$

\begin{equation}\label{eq:functional-b}
V_{\vec{q},\vec{k}} \equiv  S_q S_k S_{|\vec{q}-\vec{k}|} {\rho \over {q^4}}
\left[ {\vec{q}} \cdot
\vec{k}\,{c_k} +\vec{q} \cdot
(\vec{q}-\vec{k})\,{{c_{|\vec{q}-\vec{k}|}} }
 \right]^2
\,.
\end{equation}
The two quantities $\Omega_{q}$ and $\nu_{q}$ are respectively
the characteristic frequency of the phonon-type motions of the
fluid, and a term that describes instantaneous damping arising from
the fast contribute to the memory function. They are defined as

\begin{eqnarray}
        \Omega_{q}&=&\frac{q^{2}k_{B}T}{mS(q)} \nonumber \\
        \nu_{q}&=&\nu_{1}q^{2} \nonumber
\end{eqnarray}
and $\nu_{q}=1$ in our calculations.
In the long time limit $t\rightarrow \infty$, the density correlators
$\phi_q(t)$ tend to a value

\begin{equation}
f_q={{\langle \rho^*_q(0)\rho_q(\infty)\rangle} \over {\langle
|\rho_q(0)|^2\rangle}}
\end{equation}
the non-ergodicity factor, or Debye-Waller factor.
The MCT equations (\ref{eq:MCT}) in the static limit give rise to the
bifurcation relation

\begin{equation}
\frac{f_{q}}{1- f_{q}} = \frac{1}{2} \int \! \! \frac{d^{3}k}{(2\pi)^{3}}
\,\,
 V( {\bf q},{\bf k})f_{k}f_{|{\bf q}-{\bf k}|}
\label{eq:mctstat}
\end{equation}
It is clear that $f_q=0$ is a solution of equations~(\ref{eq:mctstat})
and it corresponds to an ergodic state of the system in which the
correlations 
decay for long time.
The correlators tend instead to a finite value different from zero
if the system is kinetically arrested.
This loss of ergodicity for $\phi$ is interpreted
as the transition to a kinetic glassy state.
Therefore for some critical values of the thermodynamic parameters,
density and 
temperature in our case, bifurcations of the solutions of the
asymptotic equations appear that produce non-zero solutions.
The bifurcations can be multiple, up to the number
of control parameters of the system. Thus,
when a bifurcation gives rise to more than two solutions of equations
(\ref{eq:mctstat}), there will exist multiple solutions
with finite non-ergodicity factors. In these cases one speaks of
$A_k$-type bifurcations, with $k=2,3,4,\dots$
In this case, MCT predicts that
only the state corresponding to the largest value of $f_q$ is a stable
solution of the equations \cite{Goetze91}.

\section{Phase diagram} %
\label{sec:phase}       %

\begin{figure}
\begin{center}
\mbox{\psfig{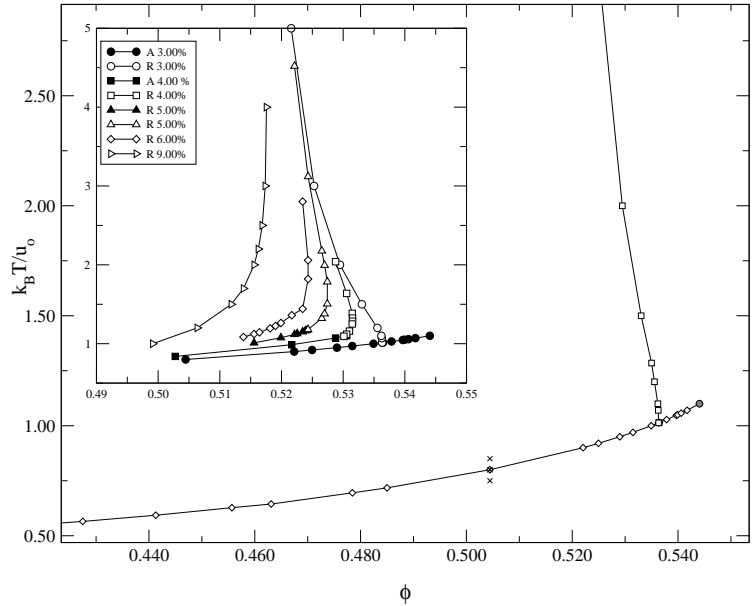}}
\end{center}
\caption{The phase diagram of a square-well system in the PY approximation
for
a square well relative width $\epsilon=0.03$. The inset shows the complete
phase diagram, where the AGL and the RGL are given for various values of
$\epsilon$. From \cite{Dawson00}.}
\label{fig:fig3}
\end{figure}

The phase diagram resulting from the solution of the MCT equations in
the large time limit is shown in figure~\ref{fig:fig3}.  We report the
liquid-glass transition lines for some values of the width parameter
$\epsilon$.  At this point it is worth reviewing the points made
earlier about the experimental observations in the introduction.
Indeed, the calculated phase diagram exhibits the phenomena of points
(i) and (ii), and in later sections we shall show that all the other
points are satisfactorily described. We point out the salient features
of the phase diagram below.

(i) For high enough temperatures the repulsive glass lines, RGL, tend to
converge to a value approaching the critical volume fraction for
hard-sphere systems $\phi_c \approx 0.516$, corresponding to the fact
that for large $T$ the short-range part of the inter-particle
potential weakly affects the dynamics. In this case the microscopic
effect described by the MCT is the excluded volume cage effect, the
impossibility of the molecules to move freely due to the presence of
the neighbouring particles.

(ii) On lowering the temperature a liquid stabilising effect due to
the attractive forces sets in, so that the liquid-glass lines tend to
extend to larger values of the volume fraction. This behaviour gives
also rise to the possibility of the glass melting when lowering the
temperature.

(iii) The reentrant behaviour characterises the interplay of the
repulsive and the attractive inter-particle forces; the system can
become a glass both when lowering or raising the temperature.

(iv) For even lower values of the temperature another phase line, the
AGL, appears and it is almost parallel to the $\phi$ axis. In this
case the particle motions are difficult since, due to the attractive
interactions, the particle tend to form bonds.

(v) The two sections of the phase lines, RGL and AGL, intersect at a
finite angle for small values of the parameter $\epsilon$.  The AGL
extends for higher volume fractions beyond the crossing point and
finally ends in a point.

(vi) The points on the glass lines we have discussed so far all
correspond to bifurcations of type $A_2$, while the end point of the
AGL relates to a higher order bifurcation $A_3$.

\begin{figure}
\begin{center}
\mbox{\psfig{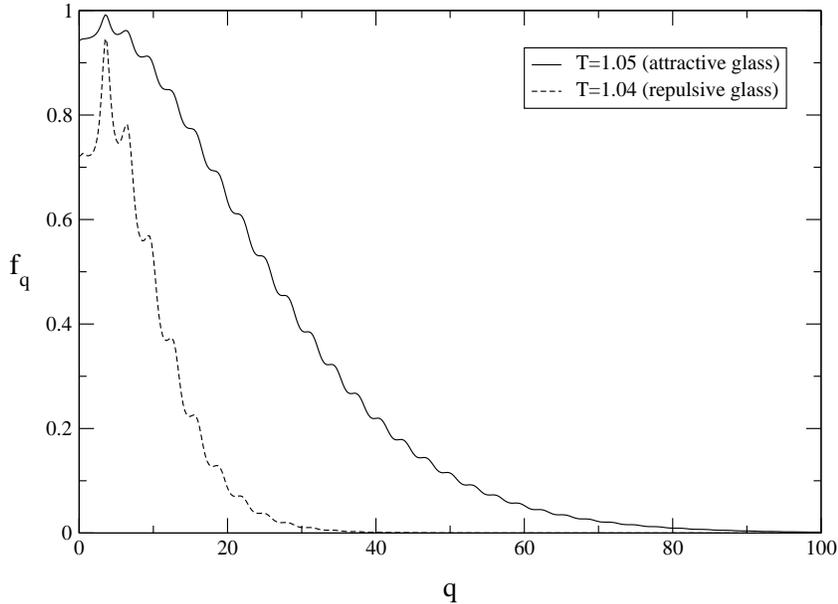}}
\end{center}
\caption{The typical non-ergodicity factors $f_q$
crossing the glass-glass transition line for $\epsilon=0.03$ and
$\phi=0.544052$.}
\label{fig:fig4}
\end{figure}

(vii) The extension of the AGL that develops in the glass region leads
to the coexistence of two different types of glasses.
In order to represent the differences
between the two glasses we present in figure~\ref{fig:fig4} the shape
of the non-ergodicity factor $f_q$ crossing the glass-glass
transition. The results are now consistent with the experimental
observation (iii) in the introduction. The width of the $f_q$ is
related to the localisation length of the particles in the glass.  If
the width is larger the particles are more localised. It is evident
that the colloidal particles are more localised in the attractive
glass then in the repulsive one. It is also possible to show that the
localisation of the length in the repulsive glass remains more or less
unchanged decreasing the temperature, whereas for the attractive one
the particles become more localised. This seems to confirm the idea
that the repulsive glass is dominated by the so-called cage effect:
when the system gets to high packing fraction the particles start to
be blocked by their neighbours, at certain critical packing fraction
$\phi_c$ each particle is trapped in a cage formed by the surrounding
particles. Indeed this is a purely geometrical effect and it should
not depend on temperature.

(viii) On increasing the relative square well width
parameter $\epsilon$ the glass-glass line tends to shrink and finally
coincides with the point of intersection of the RGL and AGL when
$\epsilon \approx 0.04$.  This end-point is a higher bifurcation point
of type $A_4$.  We note that there is no difference in the structure
factor across this curve. Only the late stage dynamics, reflected in
the values of the non-ergodicity factors, may be used to differentiate
between these glasses.

\section{The intermediate scattering function}  %
\label{sec:dyn}                                %

\begin{figure}
\begin{center}
\mbox{\psfig{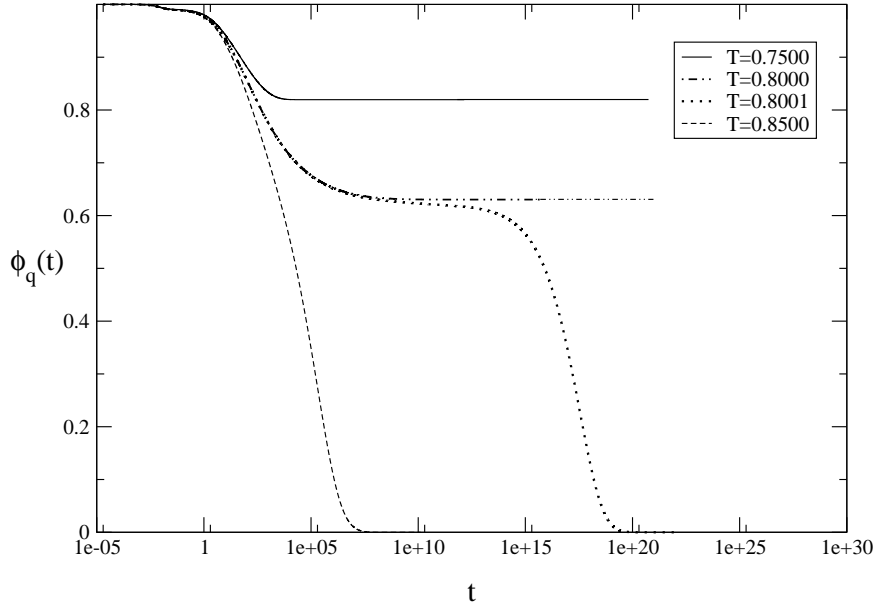}}
\end{center}
\caption{Time dependence of the density correlators for $q=21.75$,
$\epsilon=0.03$, $\phi=0.504449$ and different temperatures.}
\label{fig:fig5}
\end{figure}

We have studied the dynamical behaviour of the system in different
regions of the phase diagram, in particular in the proximity of the
higher order bifurcation points $A_3$ and $A_4$.  We will mainly
comment on the case where the well width parameter is
$\epsilon=0.03$. The wave vectors are given in units of the
particles diameter. In figure~\ref{fig:fig5} we present different
intermediate scattering functions for $q=21.75$, obtained by
solving MCT equations, approaching the glass transition at
$\phi=0.504449$ and $T_c \simeq 0.8000$.  The relevant cut across the
attractive glass curve is shown in figure~\ref{fig:fig3} using crosses, and
a
transition point. For temperatures well above the glass transition,
i.e. $T=0.85$, $\phi_q(t)$ presents the typical liquid behaviour:
after a first short-time decaying due to microscopic dynamics, the
correlation function starts to relax toward ergodicity and no sign of
a critical slowing down is evident. When the system becomes close
enough to the transition temperature ($T=0.80010$ in
figure~\ref{fig:fig5}) the system starts to freeze and the typical two
step relaxation scenario starts to be evident. We note that we are here
crossing that transition curve which we believe corresponds to the
process of gelation at high density.  We see, therefore, that we may
expect the characteristic behaviour of a glass transition in the slow
dynamics near gelation at high density.
Once the system is below the transition temperature (the case $T=0.80000$
in figure~\ref{fig:fig5}), $\phi_q(t)$ corresponding to the gel does not
relax to zero anymore, the system is non-ergodic. In this situation
there is only part of the $\beta$-relaxation toward a non-zero constant
value,
i.e. the non-ergodicity parameter $f_q \neq 0$.  Perhaps it is useful
now to discuss some of the MCT predictions for the asymptotic behaviour
of the intermediate scattering function (see for example
\cite{Goetze91} and references therein). In the time region where
$t_0\ll t \ll \tau_\alpha$ the so-called factorisation theorem holds
\begin{equation}
\label{facto}
\phi_q(t)=f^c(q)+h_qG(t)
\end{equation}
$t_0$ being a typical macroscopic time, $\tau_\alpha$ the $\alpha$
relaxation time scale and $f^c_q$ the critical non
ergodicity parameter. Equation~\ref{facto} implies a universal behaviour in
which wave-vectors and time factorize. $G(t)$ is a scaling function
which describes the whole relaxation pattern.  Near the glass
transition the behaviour of the function $G(t)$ may be calculated
analytically in the proximity of the $\alpha$ and $\beta$ relaxation.
It can be shown that the asymptotic behaviour can be expressed in
terms of a rescaled time $\tau=t/t_\sigma$ where $t_\sigma$ is a time
scale that depends crucially on the distance to the transition,
i.e. on the separation parameter $\sigma$ \cite{Goetze91}. Close to
the glass transition, for $T>T_c$, $G(\tau)$ is given by

\begin{eqnarray}
\label{asint}
 G(\tau \ll 1) \sim 1/\tau^a \\
 G(\tau \gg 1) \sim -\tau^b
\end{eqnarray}
where the two exponents are related to the exponent parameter
$\lambda$, which is obtained from a stability matrix \cite{Goetze91},
via the relation
$\lambda=\Gamma^2(1-a)/\Gamma(1-2a)=\Gamma^2(1+b)/\Gamma(1+2b)$,
which implies $0<a<1/2$ and $0<b<1$.
The scaling behaviour of the correlation function changes in the
proximity of higher bifurcation points. In particular it is possible to
show that near an $A_3$ point $G(t) \sim 1/\ln(t/t_1)$ whereas near
an $A_4$ it goes like $G(t) \sim 1/\ln^2(t/t_1)$, where $t_1$ is a
microscopic time scale~\cite{Sjoegren91}. This is quite a remarkable
behaviour
and represents a rather strong prediction of the theory. There have been
reports of such behaviour in some system \cite{Mallamace00,Sjoegren91}, and
it will be interesting to see
if this phenomenon is discovered in a variety of different colloidal
systems.

\section{Mechanical properties}  %
\label{sec:mech}                 %

Two particularly interesting quantities to investigate in the vicinity
of the glass transition are the shear viscosity and the elastic shear
modulus. Those two quantities can be measured in real systems and such
measurements can provide a good insight in the glass transition
phenomena and its relations with MCT. The complex
shear viscosity $\eta^*(\phi,T,\omega)$ can be calculated in terms of
the normalised time correlation function of the density fluctuations
$\phi_k(t)$ \cite{Bengtzelius84,Sjoegren91}
\begin{equation}
\label{eta}
\eta^*(\phi,T,\omega)={{k_B T}\over{60~\pi^2}}
{\int_0^\infty} dt e^{i \omega t}
{\int_0^\infty}dk k^4
{\left[{{dlnS_k}\over{dk}}\phi_k (t)\right]}^2
\end{equation}
and is related to the complex shear modulus $G^*(\phi,T,\omega)$
by
\begin{equation}
\label{shear}
G^*(\phi,T,\omega)= i~\omega~\eta^*(\phi,T,\omega).
\end{equation} 
It is possible to take the $\omega \rightarrow 0$ limit in equation
\ref{eta}
\begin{equation}
\label{eta0}
\eta^*(\phi,T,\omega=0)={{k_B T}\over{60~\pi^2}}
{\int_0^\infty}dk k^4
{\left[{{dlnS_k}\over{dk}} f_k \right]}^2
\end{equation}
and this gives the static shear viscosity.
We note that this formula, in that it depends on the square of the
correlation function, is different in structure from that used by Weitz
and co-workers, where the stress-relaxation function is linearly related
to the 
density correlation function~\cite{weitz}. This difference would lead,
in principle, to substantially different predictions for the loss modulus,
and in particular around its minimum. It seems likely that the idea
the stress modulus is linear in the correlation function is more suited
to visco-elastic properties dominated by large domains, and their changing
surface area under shear~\cite{onuki}, whilst the
dependence on the square of the correlation function is more suited to dense
fluids where the properties are dominated by loss of mobility due to
caging effects.
\begin{figure}
\begin{center}
\mbox{\psfig{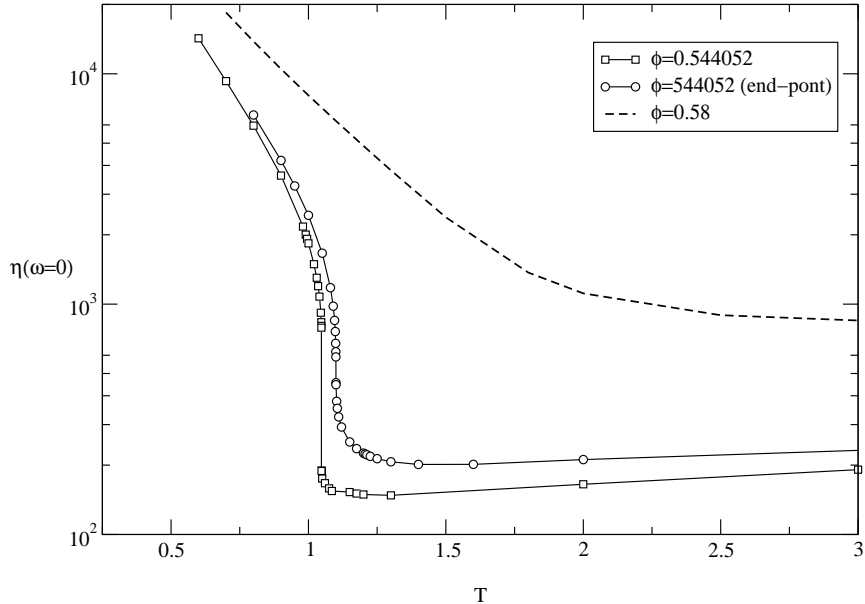}}
\end{center}
\caption{The static shear moduli on crossing the glass-glass
transition line for $\epsilon=0.03$
and close to the $A_3$ bifurcation point.}
\label{fig:fig6}
\end{figure}
All the quantities in the equations (\ref{eta}), (\ref{shear}) and
(\ref{eta0}) can be evaluated for the present system, and therefore it
is possible to use the numerical solution of equation (\ref{eq:MCT})
in order to solve them. This has been accomplished performing the
integrals with standard numerical integration over $2000$ $q$-vector
varying between $0$ and $500$.  In figure~\ref{fig:fig6} the static
shear modulus has been represented for three different packing
fractions ($\phi=0.539672$, $\phi=0.544052$ and $\phi=0.580000$) as a
function of the temperature. These three cases correspond to crossing
the glass-glass line, crossing the $A_3$ point and a density
where the repulsive and attractive glasses have become
indistinguishable. In the first case it is clearly possible to
distinguish between the two glasses. For low temperatures there is
strong dependence of the elastic viscosity on the temperature, in
particular the system becomes more and more rigid on decreasing the
temperature. When the system crosses the glass-glass transition there
is a discontinuity in the elastic response which clearly indicates
that the structure is changed.  Increasing further the temperature the
elastic behaviour does not change so much anymore and for large
temperature the system behaves like a hard spheres suspension. In this
case the glass is originated by the cage effect and consequently the
particle are forced to move inside a fixed volume that does not change
with temperature. It is then evident how the differences between the
two glasses are manifested by mechanical properties. We have noted that
the difference in the shear modulus as the $A_3$ point is approached is
described a by a power law, and it is also possible to show that in
this regime the differences in mechanical properties of the two
systems are due to the different shape of the non-ergodicity parameter
$f_q$, and hence the long time residual motions in the gel, and not to
the contribution of the equilibrium structure factor~\cite{mechpaper}.
\begin{figure}
\begin{center}
\mbox{\psfig{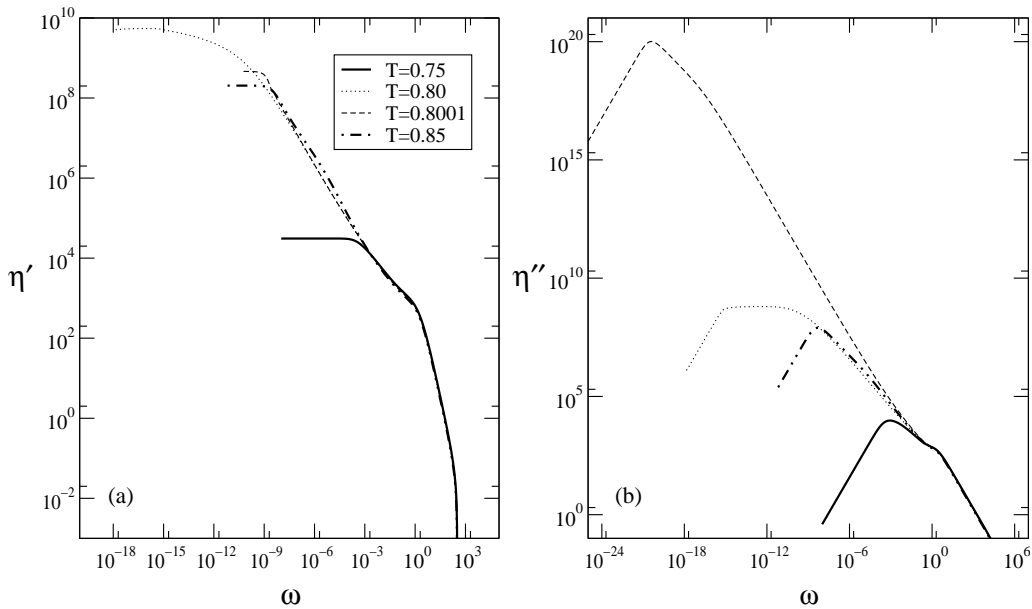}}
\end{center}
\caption{Relaxation of the shear viscosity on crossing the attractive
transition line for $\epsilon=0.03$ and  $\phi=0.504449$.}
\label{fig:fig7}
\end{figure}
We have performed this calculation also for the dynamical quantities
defined by (\ref{eta}) and (\ref{shear}). It is evident that, if the
system is in the glassy phase we have for the long time behaviour of
the density correlators $\phi_q(t\rightarrow \infty)=f_q\neq0$, and
consequently the integral in (\ref{eta}) is unbounded. Thus the
solution posses a zero frequency pole.  In order to obtain convergent
solutions for a glass we have replaced $\phi_q(t)$ with
$\hat\phi_q(t)=\phi_q(t)-f_q$ which decays to zero at infinite time.
In figure~\ref{fig:fig7}a and figure~\ref{fig:fig7}b the real and the
imaginary part of the complex viscosity, $\eta'(\omega)$ and
$\eta''(\omega)$, are presented for a fixed volume fraction,
$\phi=0.504449$, for different temperatures, approaching and crossing
the glass transition.  In figures~\ref{fig:fig8}a and \ref{fig:fig8}b
$G'(\omega)$ and $G''(\omega)$ are shown for the same points as
figure~\ref{fig:fig7}. Since figures~\ref{fig:fig7}  and \ref{fig:fig8}
contain the same information
we discuss only figure~\ref{fig:fig8}.  $G'(\omega)$ represents the elastic
response of the material: the higher its value the stiffer the
material at that frequency scale. The imaginary part of the shear
modulus describes the viscous behaviour of the system and so the
dissipation. Both quantities can be easily measured experimentally,
for example as the response of a sample to small oscillatory shears
which weakly perturb the system~\cite{weitz}.
Approaching the glass transition we note that the $G'(\omega)$ begins
to develop a shoulder at low frequency. This effect is due to the
slowing down of the dynamics of the system, in other words the
formation of the plateau in the $\phi_q(t)$ is responsible for the
formation of a region where $G'(\omega)$ varies only slightly with
frequency. The range of such a region tends to increase approaching
$T_g$. Indeed a similar behaviour has been observed experimentally in
measurements of linear viscoelasticity in a colloidal suspension with
hard sphere interaction~\cite{weitz}. If the system is in
a glassy state the value of the real part of the static shear modulus
$G'(\omega=0)$ is finite, indicating that the system is solid and
consequently it presents an elastic behaviour. In the liquid phase,
however, the system does not show any elastic behaviour so the
$G'(\omega=0)$ tends to zero. It is then clear that at the glass
transition the system abruptly changes its behaviour, presenting a
singularity in the static shear modulus.  In figure~\ref{fig:fig8}b the
behaviour of the imaginary part of the shear modulus is
represented.  At high frequency the curves for different temperature
are the same, showing the microscopic dynamics which is the same for
all the temperatures.  Approaching the glass transition a second
maximum starts to emerge.  Such a maximum represents the
$\alpha$-relaxation and it moves toward low frequency on decreasing
the distance to the glass transition temperature $T_g$. The minimum of
$G''(\omega)$ corresponds to the plateau region in time, and the power
law behavior in frequency on both sides has been already observed in
hard spheres systems~\cite{weitz}.

\begin{figure}
\begin{center}
\mbox{\psfig{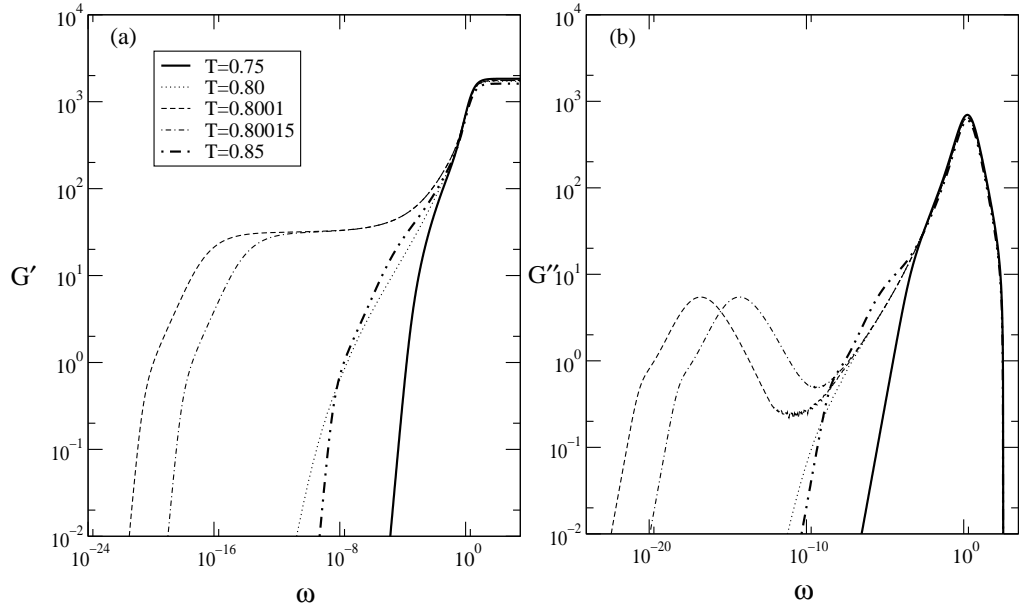}}
\end{center}
\caption{Relaxation of the shear modulus on crossing the attractive
transition line for $\epsilon=0.03$ and $\phi=0.504449$.}
\label{fig:fig8}
\end{figure}

\section{Conclusions}            %
\label{sec:conc}                 %

In this paper we have outlined some of the experimental observations
that have been made of particle gels in colloidal systems where we
know there to be a strong short-ranged potential. We have shown that
all of these features may be interpreted within the paradigm that such
particle gels represent a new type of attractive glass. However,
having made this correspondence, we then find a number of predictions
of new phenomena that have yet to be confirmed experimentally, or are
just being so. These include the re-entrant behaviour for the phase
diagram, the logarithmic dynamic behaviour near this re-entrant regime,
and in the general the characteristic alpha relaxational behaviour
near gelation.  We have calculated the mechanical manifestations of
these phenomena also.  Thus we have shown how the glass-glass
transition may be studied using the shear modulus, and that the
particle gel must be expected to be much stiffer than the colloidal
glass, but that the differences between them vanishes near the $A_3$
point.  We have also studied, to our knowledge for the first time, the
frequency dependent modulus and loss modulus for these systems, using
the paradigm that the system is undergoing a glass transition. MCT type
treatments seem to be suited to study colloidal systems of this type
because, at least of the repulsive system, the glass transition seems
to lack some of the fluctuations of molecular liquids. We also can
calculate phase diagrams, and mechanical and scattering properties
from a single theory. We expect that, in
future, dense particle gels will be better fitted to the theory
present here, than previous phenomenological treatments.

There remains much to be done in many directions. It now begins to
seems likely that the glass paradigm is suitable for interpretation of
particle gelation. However we must await further detailed
experimental work to see just how extensive the agreement with MCT
type treatments will be.  We may also speculate that this type of idea
will have much broader significance that particle gels. There is every
reason to hope that polymer gelation and protein gelation may also
find a description in similar terms.  Certainly this is a rich area
for future exploration.

\ack
{P.T. thanks Sow-Hsin Chen, to whom this paper is dedicated, for the
invaluable example and the scientific collaboration lasting for more
than three decades, and still active. K.D. and F.S. also thanks
Sow-Hsin Chen for many stimulating conversations over a period of
years, and  spanning a
number of complex fluid fields. We all appreciate his continued example
in commitment to the scientific endeavour after so many years.
The authors
also acknowledge the collaboration with the group of
Prof. Wolfgang G\"oetze,
together which part of the work reported here was peformed. The
work in Rome is supported by PRIN-2000-MURST and PRA-HOP-INFM, and the
work both in Rome and in Dublin is supported by COST P1.}

\end{document}